\documentclass[twocolumn,prb,showpacs,superscriptaddress]{revtex4}
\usepackage{graphicx}
\usepackage{dcolumn}
\usepackage{bm}
\usepackage{amsmath}
\usepackage{nicefrac}
\RequirePackage{xspace}
\newcommand{\ie}{\textit{i.e.}\xspace}

\newcommand{\units}[1]{\ensuremath{\,\mathrm{#1}}}

\newcommand{\OPI}[1][20]{\ensuremath{#1\times(0\text{-}\pi\text{-})}\xspace}

\begin{document}
\title{Interference patterns of multifacet \OPI[20] Josephson junctions with ferromagnetic barrier}
\author{S.~Scharinger}
\author{C.~G\"{u}rlich}
\affiliation{%
  Physikalisches Institut -- Experimentalphysik II and Center for Collective Quantum Phenomena,
  Universit\"at T\"ubingen, Auf der Morgenstelle 14,
  D-72076, T\"ubingen, Germany
}

\author{R.G.~Mints}
\affiliation{The Raymond and Beverly Sackler School of Physics and Astronomy, Tel
Aviv University, Tel Aviv 69978, Israel}

\author{M.~Weides}
\affiliation{%
  Institute of Solid State Research and JARA-Fundamentals of Future Information Technology, 
  Research Center J\"{u}lich, 
  D-52425 J\"{u}lich, Germany
\footnote{Current address: Department of Physics, University of California, Santa Barbara,CA 93106, USA}
}

\author{H.~Kohlstedt}
\affiliation{%
  Nanoelektronik, Technische Fakult\"{a}t, Christian-Albrechts-Universit\"{a}t zu
Kiel, D-24143 Kiel, Germany
}

\author{E.~Goldobin}
\author{D.~Koelle}
\author{R.~Kleiner}
\affiliation{%
  Physikalisches Institut -- Experimentalphysik II and Center for Collective Quantum Phenomena,
  Universit\"at T\"ubingen, Auf der Morgenstelle 14,
  D-72076, T\"ubingen, Germany
}

\date{\today}

\begin{abstract}

We have realized multifacet Josephson junctions with periodically alternating critical current density (MJJs) using superconductor-insulator-ferromagnet-superconductor heterostructures. We show that anomalous features of critical current vs. applied magnetic field,  observed also for other types of MJJs, are caused by a non-uniform flux density (parallel to the barrier) resulting from screening currents in the electrodes in the presence of a (parasitic) off-plane field component. 
\end{abstract}


\maketitle

\section{Introduction}
\label{Sec:Intro}

Multifacet Josephson junctions (MJJs), with critical current density $j_c$
alternating many times between positive $j_c^0$ and negative $j_c^\pi \approx -j_c^0$
values along the junction were intensively treated during the last years.
Initial studies were motivated by the discovery of high-$T_c$ superconductors
(cuprates) with $d$-wave order parameter symmetry \cite{Tsuei00}. In thin
film cuprate grain boundary junctions with a $45^\circ $ misalignment angle
between the two electrodes the current $j_c$ is changing its sign randomly
on a scale of a facet. A more controlled MJJ can be produced
in the form of Nb/cuprate ramp junctions where the barrier forms a zigzag
line parallel to the cuprate crystallographic axes
\cite{Smilde02,Ariando05,Guerlich09}.

A more general interest in MJJs came from the possibility to have a \emph{ground state} where the Josephson
phase has a value $\varphi$ different from both 0 and $\pi$ \cite{Buzdin03,Goldobin07a,Buzdin08,Zazunov09,Gumann09}, or to realize tunable plasmonic crystals\cite{Susanto05}.
If  a $\varphi$ junction is long compared to the Josephson length $\lambda_J$ it
may carry \emph{mobile} fractional flux quanta (splintered vortices) \cite
{Mints98,Mints01,Mints02,Moshe07b}. 
A promising option to produce an MJJ is given by the
Superconductor-Insulator-Ferromagnet-Superconductor (SIFS) technology
\cite{Kontos02,Weides06a,Weides07a}, providing exponentially
low damping at low temperatures and a high topological flexibility in arranging
the 0 and $\pi$ segments. Recent imaging of the supercurrent distribution 
showed that in SIFS \cite{Guerlich09b} $j_c$ is more homogeneous 
than in Nb/cuprate zigzag MJJs \cite{Guerlich09}.

The dependence of the critical current $I_c$ on the applied field $H$ for an MJJ
is very different from usual Josephson junctions. If $j_c^\pi = -j_c^0$ and the
facets have equal size, $I_c = 0$ in zero field but becomes large when the
field causes constructive interference of the supercurrents flowing through the 0 and
$\pi$ segments. Standard calculations of $I_c(H)$ for a junction of length $L$
and width $W$, assuming that the flux density $B$ is homogeneous in the tunneling barrier and
$L,W\lesssim 4\lambda_J$, show that the main peaks occur if the flux per facet
is $\Phi_0/2$ ($\Phi_0$ is the flux quantum), resulting in the maximum current $I_c = 2LWj_c^0/\pi$. The other
maxima in $I_c(H)$ are much lower and symmetric relative to the main $I_c$
maxima \cite{Smilde02}.

Suppression of $I_c(0)$ and appearance of high-field main maxima were clearly
observed for grain boundary MJJs in a field perpendicular to the substrate plane \cite{Copetti95,Mannhart96,Hilgenkamp02}. Here, randomness in $j_c^0, j_c^\pi$ and facet sizes prevent a close comparison to the ``ideal'' theoretical $I_c(H)$. However, differences between experiment and theory also appear for the more controllable zigzag MJJs. In particular, the secondary maxima between the two main maxima almost vanish in experiment and are strongly enhanced outside the main peaks \cite{Smilde02,Ariando05,Guerlich09}. These features cannot be reproduced in calculations by simply taking into account nonuniformity of the junctions.

\begin{figure}[tb]
  \begin{center}
    \includegraphics{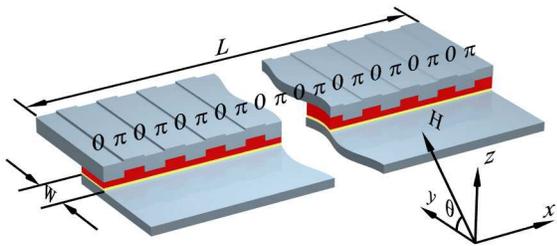}
  \end{center}
  \caption{(Color online). Sketch of a SIFS MJJ with 20 facets.}
  \label{fig:sketch}
\end{figure}

A major step towards understanding $I_c(H)$ was done in the framework of nonlocal
electrodynamics of MJJs \cite{Moshe09}. A universal solution has been found for $\phi (x)$ in the case when the electrodes are formed by an ultrathin film and the applied
field is perpendicular to them (the definition of the coordinates is shown in Fig.~\ref{fig:sketch}). In this geometry, the flux density $B_z(x)$ is strongly enhanced compared to $\mu_0H_z$, has a maximum in the center of the junction and decays to zero towards the edges, see Fig.~2(a) in Ref.~\onlinecite{Moshe09} (the quantity $\phi_0^\prime(Y)$ plotted there is proportional to the flux density along the junction, i. e. $B_z(x)$ in our notation). As a result the main maxima of $I_c$ are suppressed compared to the case of uniform $B_z$ while the maxima following the main maximum are strongly enhanced, see
Fig.~4(b) of Ref.~\onlinecite{Moshe09}.

In this paper we report the results of our study of $I_c(H)$ dependence of rectangular $\mathrm{Nb|Al_2O_3|Ni_{0.6}Cu_{0.4}|Nb}$ SIFS MJJ structures with length $L=200 \units{\mu m}$ and width $W=10 \units{\mu m}$. Sections, consisting of 5 \units{\mu m} long 0-$\pi$ segments are repeated $N=20$ times, as sketched in Fig.~\ref{fig:sketch}. The total length $L\approx 3 \lambda_J$, {\it i.e.}, these MJJs can be treated as short junctions.

\section{Theory}
\label{Sec:Theory}

In a short MJJ, see Fig.~\ref{fig:sketch}, the flux generated by tunneling currents is negligible. If
$B_y$ and $j_c$ do not depend on $y$, $I_c$ is given by
\begin{equation}
I_c= W\max_{\phi_0}{\int_{-L/2}^{L/2} \left\{ j_c(x)\sin[\phi(x,B_y)+\phi_0] \right\}}\,dx.
\label{Eq:I_c_H_calc}
\end{equation}
In what follows we consider complex interference patterns depending on the spatial distribution of flux inside the junction. Thus, to show the results of our numerical studies, we plot $I_c/I_{c0}$ as a function of $H/H_0$, where $I_{c0}=j_c^0LW$ and $H_0=\Phi_0/\mu_0 \Lambda L$. $\Lambda$ is the (average) effective magnetic junction thickness accounting for flux penetration into the superconducting electrodes and the effect of screening currents. ($\Lambda \approx 2\lambda_L$ for a junction in a bulk sample, where $\lambda_L$ is the London penetration depth); below we will also allow $\Lambda$ to be different in the 0 and $\pi$ parts.  
\par
\begin{figure}[tb]
    \includegraphics[width=\columnwidth,clip]{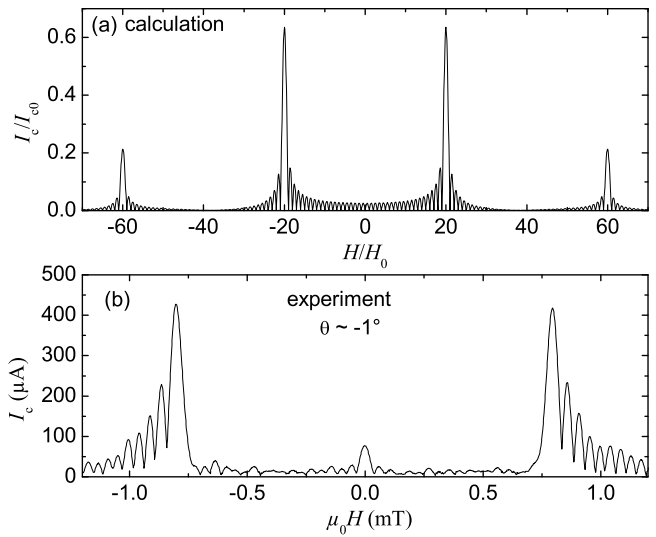}
  \caption{(Color online).
    (a) Calculated critical current of a \OPI[20] MJJ as a function of applied field, for $j_c^0=|j_c^\pi|$, facets of equal size and a homogeneous field ${\bf B}\parallel\hat{y}$. (b) A ``typical'' experimental pattern ($T=4.2\,$K), when $\bm{H}$ is misaligned relative to the substrate plane ($\theta \approx -1^\circ$).
  }
  \label{fig:1}
\end{figure}

The phase $\phi(x,B_y)$ is obtained from the equation
\begin{equation}
\frac{d\phi}{dx} = \frac{2\pi B_y(x)\Lambda}{\Phi_0},
\label{Eq:phi_calc}
\end{equation}
The calculated $I_c(H)$ for constant $\Lambda$ is shown in Fig.~\ref{fig:1}(a). The two main maxima of $I_c$ occur at $|H|/H_0 = N$. The next maxima are located at $|H|/H_0 = 3N$. 
Further, $I_c = 0$ at $|H|/H_0=2N$.

\section{Experiment}
\label{Sec:Experiment}

The current-voltage curve of the MJJs that we studied experimentally ($T\approx4.2 \units{K}$) are non-hysteretic. We determined $I_c$ by using a voltage criterion $V_\mathrm{cr} = 1 \units{\mu V}$ except for the data shown in Fig.~\ref{fig:1}(b), where we have used a different measurement setup, with $V_\mathrm{cr} = 0.2 \units{\mu V}$. Thus, we have an $I_c$ detection limit (parasitic $I_c$ background) $I_\mathrm{cr} = V_\mathrm{cr}/R$ ($R\approx 0.05 \units{\Omega}$). $I_\mathrm{cr}\approx 4 \units{\mu A}$ in Fig.~\ref{fig:1}(b) and  $\approx 20 \units{\mu A}$ in all other experiments. When the sample was aligned ``by eye'' to have $\bm{H}\parallel\hat{y}$, the $I_c(H)$ patterns (see, \textit{e.g.}, Fig.~\ref{fig:1}(b); in comparison to Fig.~\ref{fig:1}(a) the applied field scale roughly corresponds to $\pm 30H_0$) resembled the ones of zigzag junctions\cite{Smilde02,Ariando05} (where $\bm{H}\parallel\hat{z}$), but not the calculated ``ideal case'' pattern shown in Fig.~\ref{fig:1}(a).  In particular, the measured $I_c(H)$ shows a small local maximum at $H=0$ and high maxima following the main maxima.

\begin{figure}[tb]
    \includegraphics[width=\columnwidth,clip]{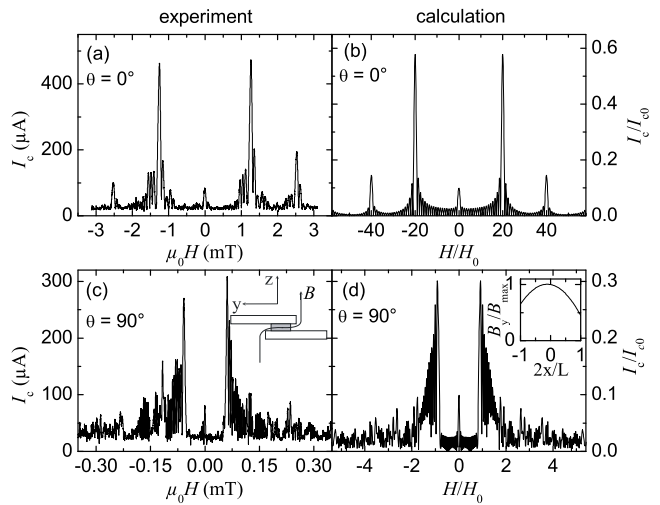}
  \caption{%
    $I_c(H)$ curves for (a,b) parallel ($\theta = 0^\circ$) and (c,d)  perpendicular ($\theta = 90^\circ$) field orientation. Inset in (c) is a sketch of $\bm{B}$ in the $(y,z)$ plane for $\theta=90^\circ$. Inset in (d) shows the profile $B_y(x)$ used to calculate $I_c(H)$ for $\theta = 90^\circ$.
  }
  \label{fig:2}
\end{figure}

The model considered in Ref.~\onlinecite{Moshe09} (ultrathin film grain boundary MJJ in the $(x,y)$ plane) cannot be applied, since (a) the geometry of SIFS MJJ is more complicated and (b) the validity conditions for the nonlocal model are not satisfied. However, for the SIFS (and also zigzag) MJJs an applied field $\bm{H}\parallel\hat{z}$ should be subject to field focusing effects, with a maximum of $B_y(x)$ in the center of the junction\cite{Monaco08a,Monaco09}. In our MJJ aligned ``by eye'' a component $H_z$ may arise from a slight out-of-plane misalignment of the applied field. This small field leads to strong modifications of $I_c(H)$. To test this idea
we measured $I_c(H)$ for different angles $\theta$ between $\bm{H}$ and the substrate plane (see Fig.~\ref{fig:sketch}), using two perpendicular coils operated in linear combination and creating $H_y=H\cos\theta$ and $H_z=H\sin\theta$.

Fig.~\ref{fig:2}(a) shows $I_c(H)$ measured at an angle, which, after having measured and analyzed the angle dependent $I_c(H)$ (see below), has been identified to be very near $\theta = 0^\circ$. The shape of this curve is much closer to the ``ideal case'', although differences occur. First, there is an $I_c$ maximum at $H=0$. This naturally arises, because, typically, $|j_c^\pi| \neq j_c^0$ and thus the cancelation of the supercurrents carried by the 0 and $\pi$ parts is incomplete. Second, an $I_c$ peak appears at $H/H_0 = 40 = 2N$. If both the 0 and $\pi$ segments had the same length and the same $\Lambda$, at this field $I_c$ should cancel within each segment. However, $\Lambda$ may slightly differ for the 0 and $\pi$ parts. Then, $I_c$ cannot cancel at the same field in both the 0 and the $\pi$ segments and a peak may appear in $I_c(H)$.
Further note that the measured $I_c(H)$ is asymmetric with respect to positive and negative $H$. This effect arises from asymmetries in the ferromagnet layer magnetization which we do not take into account to keep the discussion short.
\par
Taking asymmetries of $j_c$ and different values $\Lambda_0$ and $\Lambda_{\pi}$ of the 0 and $\pi$ parts into account, using Eq.~\eqref{Eq:I_c_H_calc} and Eq.~\eqref{Eq:phi_calc},  we calculate the $I_c(H)$ pattern shown in Fig.~\ref{fig:2}(b). 
In Fig.~\ref{fig:2}(b) we have used $\Lambda_{\pi}/\Lambda_0 = 1.38$ and $j_c^\pi/j_c^0=-0.8$ to achieve the best fit to the experimental curve.
This translates to $j_c^0 \approx 42 \units{A/cm^2}$ and $j_c^\pi \approx -34 \units{A/cm^2}$ in physical units.
These parameters are reasonable and the resulting $I_c(H)$ describes the data well. Also note that the ratios $j_c^\pi/j_c^0$ and $\Lambda_{\pi}/\Lambda_0$ can be determined independently and in a straightforward way, by analyzing the $I_c$ maxima at, respectively, $H$ = 0 and $H=20H_0$. Comparing the field axes of Figs.~\ref{fig:2}(a) and (b) we obtain $\Lambda_0 \approx 134 \units{nm}$ and $\Lambda_{\pi} \approx 186 \units{nm}$ in reasonable agreement with the value of $2\lambda_L (\lambda_L \approx 90 \units{nm}$ for Nb). Further, using
\begin{equation}
  \lambda_J^i = \sqrt{{\Phi_0}/{2\pi\mu_0 |j_c^i|d^\prime}}, \text{ with } i=0,\pi. 
  \label{Eq:lambda_J}
\end{equation}
Here  $\mu_0{d^\prime}$ is the inductance per square (with respect to in-plane currents) 
of the electrodes and $d^\prime\approx \Lambda$ for electrode thicknesses $\gtrsim \lambda_L$, we find a normalized junction length $l\equiv L/2\lambda_J^0+L/2\lambda_J^\pi\approx3.2$, {\it i.e.}, this MJJ is in the short limit.

\begin{figure}[tb]
  \begin{center}
    \includegraphics[width=\columnwidth,clip]{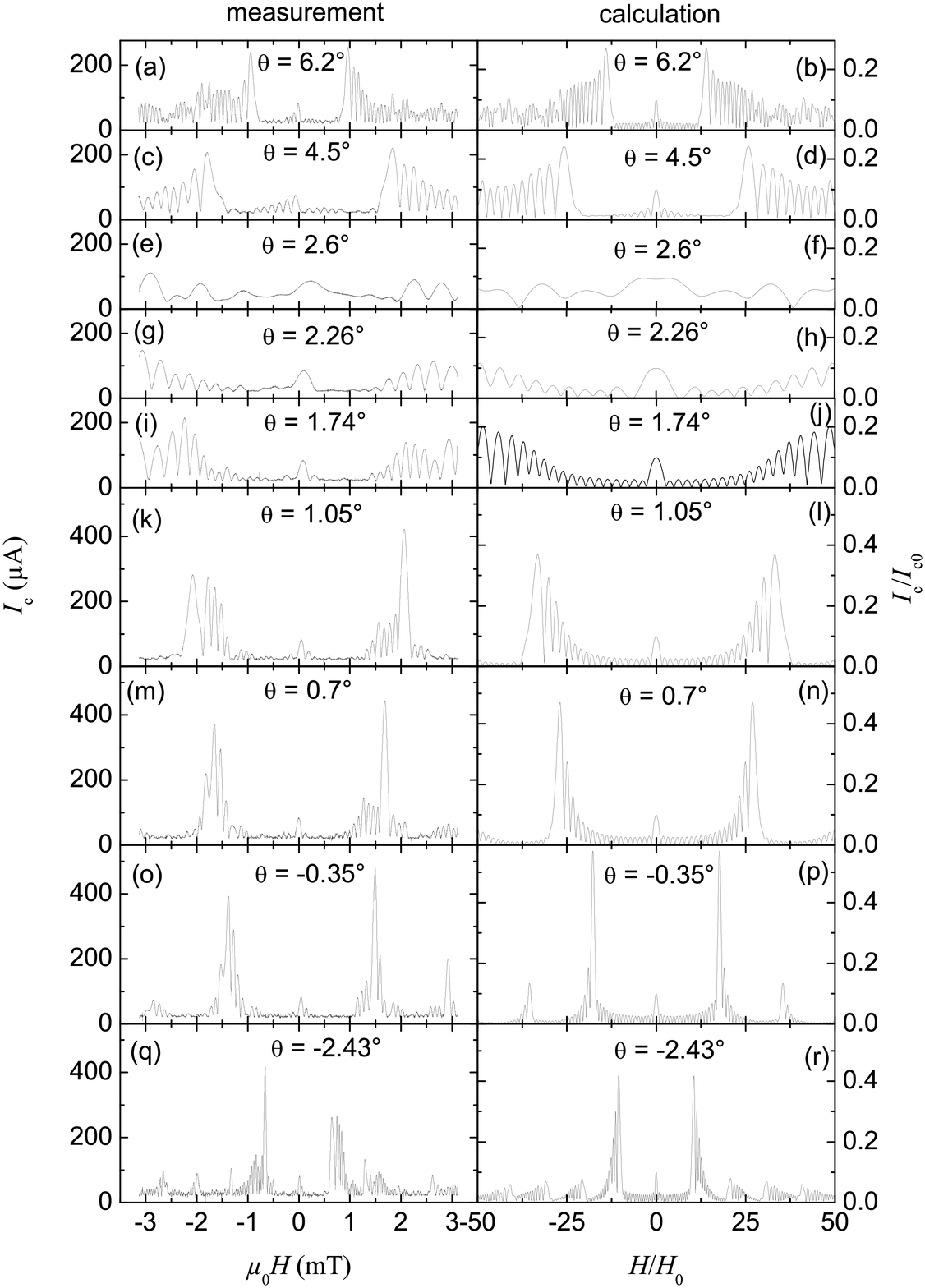}
  \end{center}
  \caption{$I_c(H)$ patterns for different misalignment angles $\theta$.}
  \label{fig:3}
\end{figure}

Having analyzed $I_c(H_y)$ and fixed the junction parameters $j_c^0$, $j_c^\pi$, $\Lambda_0$ and $\Lambda_{\pi}$ we turn to the other limit of $\bm{H}\parallel\hat{z}$, ($\theta = 90^\circ$). The corresponding measured $I_c(H)$ pattern is shown in Fig.~\ref{fig:2}(c). Its shape resembles the ones of zigzag junctions \cite{Smilde02,Ariando05,Guerlich09} or of the SIFS MJJ with $\bm{H}\parallel \hat{y}$ aligned ``by eye'', c.f.~Fig.~\ref{fig:1}(b). Note that our MJJ is sensitive to $H_z$, because screening currents flowing inside the electrodes create a non-uniform field component $B_y(x)$ inside the junction barrier, as sketched in the inset of Fig.~\ref{fig:2}(c). To calculate $I_c(H_z)$
we need to find a profile $B_y(x)$ caused by the applied field $\bm{H}\parallel \hat{z}$ and solve Eqs.~\eqref{Eq:I_c_H_calc} and ~\eqref{Eq:phi_calc} using this $B_y(x)$.
Similar to the case of a 0 junction having overlap geometry \cite{Monaco08a,Monaco09} and to the nonlocal grain boundary junction \cite{Moshe09} we expect $B_y(x)$ to have a maximum in the center of the MJJ ($x=0$). It can be approximated as \cite{Monaco08a,Monaco09,Moshe09}
$B_y(x)\propto \cos(\pi x/L)$  if the junction is ``naked'', i.e., without idle regions surrounding the Josephson barrier.
In our case there are $5 \units{\mu m}$ wide idle regions. Then, $B_y$ will be nonzero at $x=\pm L/2$. In addition, asymmetries in $B_y(x)$ may occur due to asymmetries in the junction layout. Thus, we approximate $B_y(x)$ by a Taylor expansion near $x=0$, \ie,  $B_y(x) = -f \mu_0H_z(1+a_1\xi+a_2\xi^2)$, where $\xi=2x/L$ and $-1<\xi<1$ . Solving Eq.~\eqref{Eq:I_c_H_calc}, we get the $I_c(H)$ pattern shown in Fig.~\ref{fig:2}(d). We have used $a_1 = -0.1$ and $a_2 = -0.45$ to reproduce the data. While the overall shape of $I_c(H)$ strongly depends on the value of $a_2$, the nonzero choice of $a_1$ is somewhat cosmetic and was merely necessary to suppress a ``shoulder'' in the calculated $I_c(H)$ oscillations appearing for flux values above the main $I_c$ maximum. For lower values of $H$ this parameter has only a small effect on $I_c$. The resulting field profile is shown in the inset of Fig.~\ref{fig:2}(d). 
Further, by comparing the abscissas of the calculated and experimental curves, we find a field focusing factor $f\approx 23$, which is a reasonable value for the large MJJ we study.

Having reproduced $I_c(H)$ for both $\bm{H}\parallel\hat{y}$ and $\bm{H}\parallel\hat{z}$,
we consider the general case where $\bm{H}$ is tilted by an angle $\theta$. 
No additional parameters are required for calculating $I_c(H)$. 
We obtain
\begin{equation}
B_y(\xi) = -\mu_0H[(f\sin\theta-\cos\theta) + (a_1\xi+a_2\xi^2)f\sin\theta]. \label{Eq:tilted}
\end{equation}
Similar to the geometries analyzed in Refs.~\onlinecite{Heinsohn01, Monaco09} at a ``dead angle'' $\theta_d = \arctan(1/f)$ the uniform field term vanishes and one obtains an anomalous $I_c(H)$ dependence with diverging period.
The linear and quadratic terms in $B_y(x)$ can have different signs relative to the constant term depending on $\theta$. For angles between $\theta_d$ and $0^\circ$, and for a negative coefficient $a_2$,  $B_y(x)$ peaks at the junction edges while at all other angles the maximum in $B_y$ is in the junction center.
\par
In Fig.~\ref{fig:3} we show a sequence of $I_c(H)$ patterns starting from $\theta = 6.2$ to $\theta = -2.43^\circ$ comparing the measured $I_c(H)$ with calculated curves. All data have been taken in a single cooldown. In the simulated curves the junction parameters and coefficients $a_1,a_2$ are the same as for the cases $\theta = 0°$ and $\theta=90^\circ$ discussed above. The patterns taken at $\theta = 6.2^\circ$, see Fig.~\ref{fig:3}(a) differ only marginally from the case of $\theta = 90^\circ$, shown in Fig.~\ref{fig:2}(a). The main difference is the field scale due to the $\sin\theta$-reduced $B_y(x)$ field caused by $H_z$. In particular, for the cases $\theta = 6.2^\circ$ and $\theta = 4.5^\circ$ the enhancement of the higher order $I_c$ maxima (relative to the main maximum) can nicely be seen.
When approaching $\theta_d (\approx 2.5^\circ$ for our sample), $I_c(H)$ stretches anomalously with an increased amplitude of the low-field $I_c$ oscillation and with an increased field modulation period. Further, for angles between $\theta_d$ and $0^\circ$ (where, according to Eq.~\eqref{Eq:tilted}, the field profile $B_y(x)$ reverses shape) the \textit{lower} order $I_c$ maxima become enhanced while the \textit{higher} order maxima are suppressed.
These features are well reproduced by the calculated curves $I_c(H)$.
$I_c(H)$ at $\theta = 0^\circ$ is shown in Fig.~\ref{fig:2}(a). For  negative values of $\theta$ the shape of $I_c(H)$ rapidly develops into the pattern found for $\bm{H}\parallel\hat{z}$, without anomalies at $\theta = -\theta_d$.

\begin{figure}[tb]
    \includegraphics[width=\columnwidth,clip]{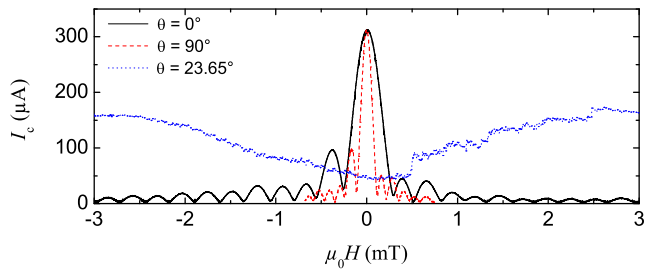}
  \caption{(Color online).
    $I_c(H)$ curves for a $50\times10  \units{\mu m^2}$ SIFS 0 junction for $\theta = 0^\circ,90^\circ$ and $23.65^\circ (\approx\theta_d)$.
  }
  \label{fig:4}
\end{figure}

A ``dead angle'' has been found for 0 junctions of various geometries \cite{Heinsohn01,Monaco09}. To further test this for SIFS junctions having large idle regions we have measured $I_c(H)$ of a $50\units{\mu m}\times 10 \units{\mu m}$ 0 coupled SIFS junction, for $\bm{H}$ applied under different angles $\theta$. Fig.~\ref{fig:4} shows the resulting curves for three values of $\theta$. For $\theta = 0^\circ$ and $90^\circ$ $I_c(H)$ is close to a Fraunhofer pattern, with a compressed field scale for the $\theta = 90^\circ$ case; by comparing the modulation periods of these two curves we find $f \approx 2.3$ and calculate $\theta_d \approx 23.5^\circ$. The dependence $I_c(H)$, measured at $\theta=23.65^\circ$ and shown in Fig.~\ref{fig:4}, indeed is by no means  Fraunhofer like.

The value of $\theta_d$ depends on the lead geometry via the field focusing factor $f$ which is large for long junctions with wide leads. While $\theta_d \approx 23.65^\circ$, found for our 0 junction, is not very critical  if the sample is aligned ``by eye'', other geometries, particularly in the context of long junction physics may have larger values of $f$ and correspondingly lower values of $\theta_d$.
\par

\section{Summary}
\label{Sec:Summary}

We have studied a \OPI[20] multifacet SIFS Josephson junction. Its $I_c(H)$ dependence, when the applied magnetic field is aligned accurately parallel to the junction plane, can be described well by the standard linear phase ansatz (constant field), taking into account $|j_c^0|\neq |j_c^\pi|$ and different values for the effective magnetic junction thickness $\Lambda$ in the 0- and $\pi$-parts.
On the other hand, a variety of $I_c(H)$ patterns can be obtained as a result of a small misalignment and field focusing. When the perpendicular field component dominates, the $I_c(H)$ patterns are similar to the ones measured for Nb/cuprate zigzag 0-$\pi$ junctions. Specific features are: (a) a suppressed amplitude of the main $I_c$ maxima and (b) a strong asymmetry between the lower order and higher order maxima.
If the field is off-plane one may meet a ``dead angle'' $\theta_d$ where the flux density $B_y$ caused by $H_y$ is cancelled by the constant part of a non-uniform $B_y(x)$ induced by  $H_z$. The effect is not restricted to multifacet SIFS junctions.
If the sample is aligned so that $\theta \approx \theta_d$, $I_c(H)$ is anomalous and may lead to erroneous conclusions about the sample quality or even the physics investigated. 

Although the MJJ investigated here has a too large $j_c$ asymmetry to from a $\varphi$ junction, the effects discussed here should be taken into account for the right interpretation of experimental results obtained for such devices in applied magnetic field. 

\acknowledgments

We acknowledge financial support of the German Israeli Foundation (Grant No. G-967-126.14/2007) and of the Deutsche Forschungsgemeinschaft (SFB/TRR 21 and project WE 4359/1-1).

\bibliography{SF,SIFS_REFs}
\end{document}